\documentclass[twoside,a4paper,11pt]{sea10}
\usepackage{graphicx}
\usepackage{hyperref}
\usepackage{movie15}
\begin{document}
\pagenumbering{arabic}
\pagestyle{myheadings}
\thispagestyle{empty}
{\flushleft\includegraphics[width=\textwidth,bb=58 650 590 680]{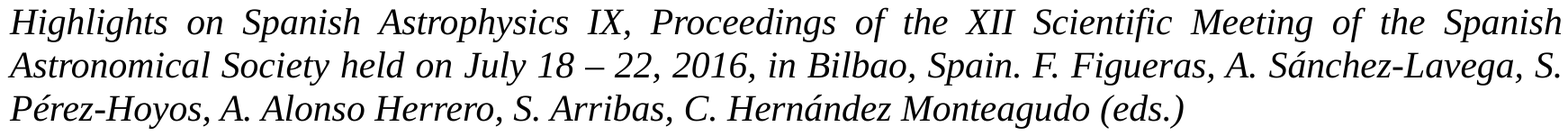}}
\vspace*{0.2cm}
\begin{flushleft}
{\bf {\LARGE
%
%%% TITLE of the paper. 
%%% TITLE of the paper. 
Early-type massive stars in Carina Nebula within the Gaia-ESO Survey. 
%
% Do not delete next few lines
}\\
\vspace*{1cm}
%
%%% Include here the LIST OF AUTHORS.
%%% Include here the LIST OF AUTHORS.
%%% Note that the last author has to be preceeded by an AND.
S.R. Berlanas$^{1,2}$,
A. Herrero$^{1,2}$, 
F. Martins$^{3}$, 
S. Sim{\'o}n-D{\'i}az$^{1,2}$,
L. Mahy$^{4}$,
R. Blomme $^{5}$,
and 
the GES WG-13
%
% Do not delete next few lines
}\\
\vspace*{0.5cm}
%
%%% AFFILIATIONS LIST.
%%% and the AFFILIATIONS LIST. Note that one affiliation per line.
%%% Add as many affiliations as necessary. 
$^{1}$
Instituto de Astrof{\'i}sica de Canarias, 38200 La Laguna, Tenerife, Spain\\
$^{2}$
Departamento de Astrof{\'i}sica, Universidad de La Laguna, 38205 La Laguna, Tenerife, Spain\\
$^{3}$
LUMP, Universit{\'e} de Montpellier, CNRS, Place Eug{\`e}ne Bataillon, F-34095 Montpellier, France\\
$^{4}$
Institut d{\ ’}Astrophysique et de G{\'e}ophysique, Universit{\'e} de Li{\`e}ge, B{\^a}t B5C, All{\'e}e du 6 Ao{\^u}t 17, 4000 Li{\`e}ge, Belgium\\
$^{5}$
Royal Observatory of Belgium, Ringlaan 3, 1180 Brussels, Belgium

%
% Do not delete next few lines
\end{flushleft}
%
% Headings
\markboth{
%%% Type the SHORT version of the paper title.
%%% Type the SHORT version of the paper title.
Early-type massive stars in Carina Nebula within the Gaia-ESO Survey
}{ % Do not delete
%
%%%  First Author \& Second Author   OR   First-author et al. 
%%%  First Author \& Second Author   OR   First-author et al. if the author list 
%%% contains three or more authors.
S.R.Berlanas et al.
% 
% Do not delete next few lines
}
\thispagestyle{empty}
\vspace*{0.4cm}
\begin{minipage}[l]{0.09\textwidth}
\ 
\end{minipage}
\begin{minipage}[r]{0.9\textwidth}
\vspace{1cm}
\section*{Abstract}{\small
%
% ABSTRACT ABSTRACT ABSTRACT
% ABSTRACT ABSTRACT ABSTRACT
%%% Type the ABSTRACT of your paper

The Gaia-ESO Survey (GES) is obtaining high quality spectra of $\sim$10$^{5}$ stars in our Galaxy, providing an homogeneous and unique overview of all the main components of the Milky Way, its formation history and the evolution of young, mature and ancient Galactic populations.\
Our group is in charge of the early-type massive stars that define the youngest population in the survey. In this contribution, we present the results of the quantitative spectroscopic analysis of O-type stars in the Carina Nebula within the Gaia-ESO Survey. For this aim, we have used FASTWIND and CMFGEN stellar atmosphere codes, providing stellar parameters for the current sample (GES data release iDR4).

%
% Do not delete next few lines
\normalsize}
\end{minipage}
%
%
%%% BODY of the paper
%%% BODY of the paper
%
\section{Introduction \label{intro}}

The Gaia-ESO Survey is covering the main components of the Milky Way using VLT-FLAMES, providing an homogeneous overview  of the kinematics and chemical composition of our Galaxy. Open Clusters are useful tools for this aim, where it is possible to study stellar populations of different ages in different evolutionary stages. Combined with Gaia astrometry, this survey will help us to understand the formation history and evolution of young, mature and ancient Galactic populations. 

The Carina Nebula Association represents a unique region to study Galactic massive stars with FLAMES. On the one hand, is one of the most massive association at a nearby distance $\sim$ 2.3 kpc  \cite{smith02} where is expected to obtain spectra for hundreds of massive stars. On the other hand, there is no systematic analysis of the stellar parameters in the association, that will be an important contribution of the Gaia-ESO Survey.\
The analysis of the Carina massive stellar population will be highly relevant for problems like the Initial Mass Function  \cite{figer05}, \cite{crowther10}, the detailed chemical composition and evolution (see \cite{ssimon10} in Orion) or the stellar multiplicity \cite{sana10}.

%Different studies have been done in hot clusters with massive stars. An example is the one done in the Arches Cluster, where was determined the upper mass limit using the IMF \cite{figer05}. Another is the determination of the chemical composition of the Orion Nebula \cite{ssimon10}, or the review of multiplicity of massive stars \cite{sana10}.
%Therefore, we will be able to do similar studies in the Carina Nebula determining the stellar evolution in the region.

\section{Gaia-ESO spectra \label{s1}}

The Gaia-ESO Survey (GES) is obtaining spectra using the Giraffe and UVES instruments, and for early type stars in Trumpler 14 cluster are employing four Giraffe setups (HR03/ 05A/ 06/ 14A) and the UVES CD3 (520-nm setting).
%In Figure~\ref{fig1} is represented the covered wavelength range and the lines available for the spectroscopic analysis in each case.
Unfortunately, the H$\gamma$ line is not covered in the Giraffe setting, which provides most of the early-type spectra. This is a problem for our spectroscopic analysis, due to its importance on the accurate gravity determination (is not affected by other broadening effects).

\begin{figure}[!h]
\center
%\par{
\includegraphics[width=8.5cm]{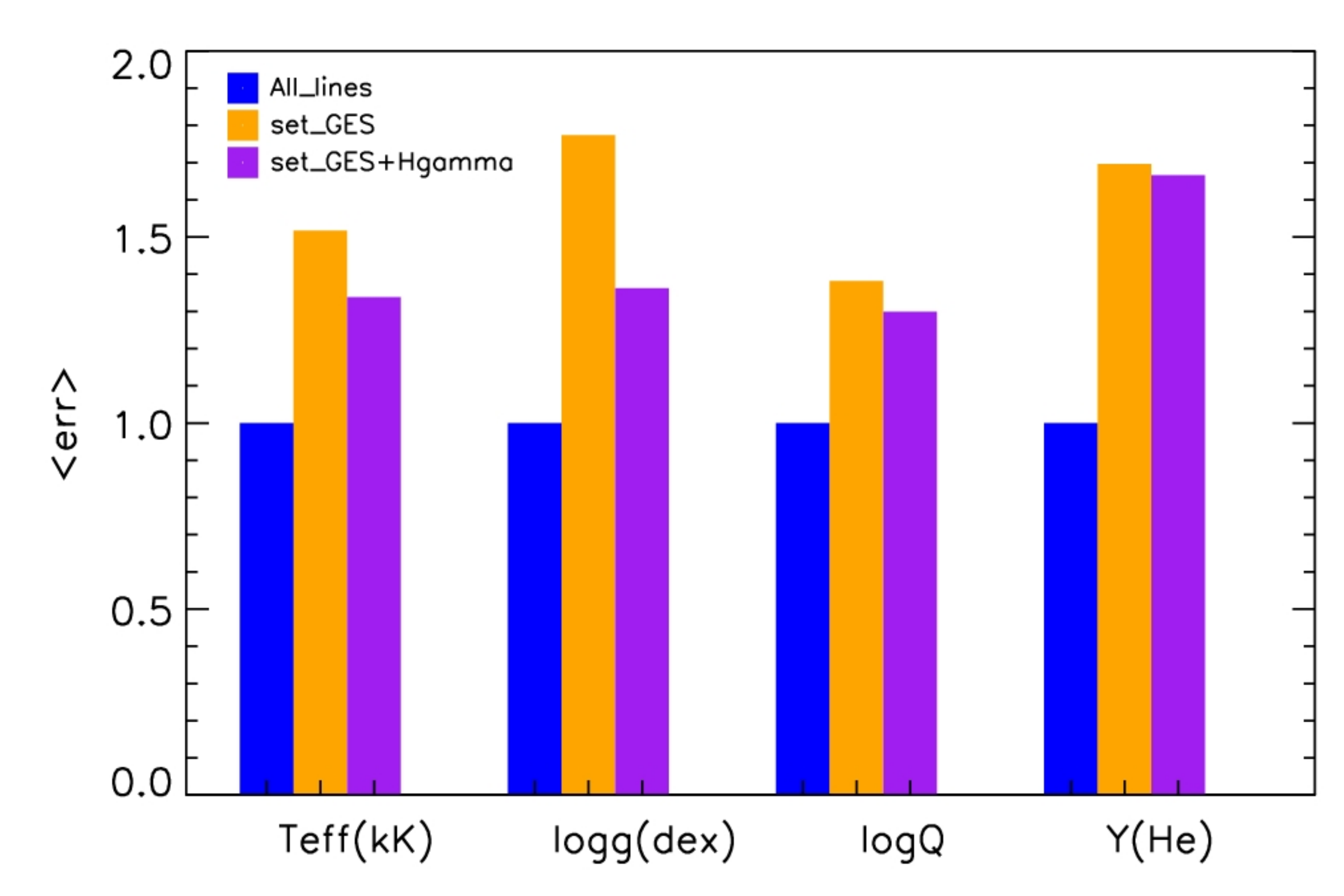}

%\par}
%\includegraphics[width=8.3cm,angle=0,clip=true]{sea_logo.pdf} 
\caption{\label{fig1} Histogram of the errors obtained for Teff, logg, logQ and He abundances using all the lines available on the IACOB-OWN spectra (in blue), using only the lines available on the GES spectra (in orange) and adding the Hgamma line to them (in purple). Values have been normalized to those obtained with the whole line set available (the blue bars).
}
\end{figure}

In order to know the real effects of the lack of this line, we decided to analyse 129 IACOB-OWN spectra, which have a full wavelength coverage in the optical range, but using different line sets. In Figure~\ref{fig1} we represents the errors obtained for the effective temperature (Teff), gravity (logg), wind strength parameter (logQ, see \cite{puls96}) and He abundances using (a) all the lines available on the IACOB-OWN spectra (in blue), (b) using only the lines available on the GES spectra (in orange) and (c) adding the H$\gamma$ line to them (in purple). It is clear that adding this line, we are improving significantly the GES O-type parameters determination. The effect is particularly relevant for gravity, and therefore the derived spectroscopic mass.
Based on these results, the GES Working Group 13 on massive stars proposed the addition of the HR4 grating to the observations, covering the H$\gamma$ line. This proposal was accepted by the GES Steering Committee and the ESO, and we have just received the new data, whose analysis is now started.

%\begin{figure}[!h]
%\center
%\par{
%\includegraphics[width=10.5cm]{range.pdf}

%\par}
%\includegraphics[width=8.3cm,angle=0,clip=true]{sea_logo.pdf} 
%\caption{\label{fig1} Wavelength range of the Giraffe and UVES instruments, where the available lines for the spectroscopic analysis are plotted.}

%\end{figure}

%\subsection{HR4 grating \label{s11}} 

\section{Spectroscopic analysis \label{s12}} 
\begin{figure}[!h]
\center
%\par{
\includegraphics[width=11.0cm, angle=180]{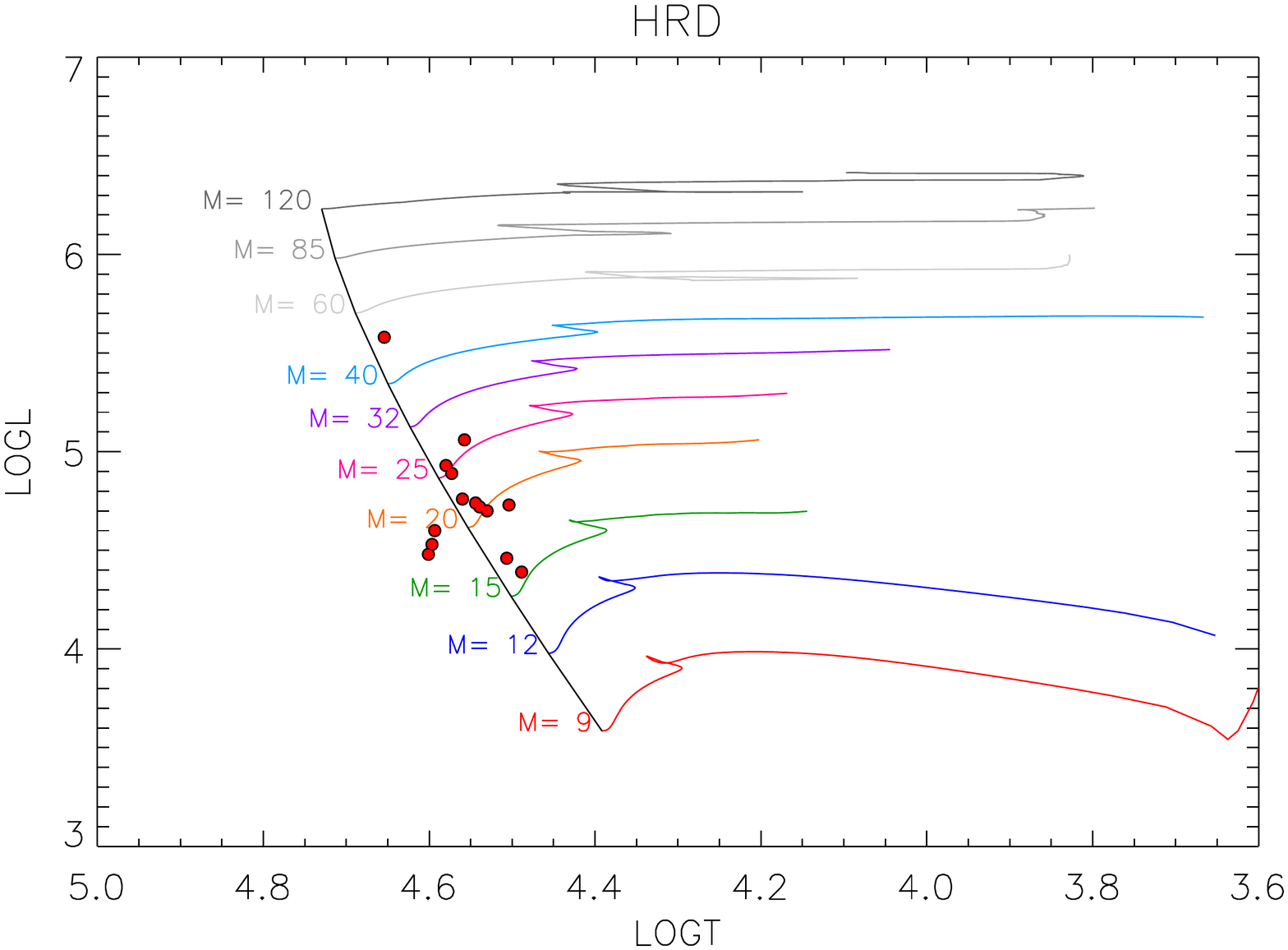}

%\par}
%\includegraphics[width=8.3cm,angle=0,clip=true]{sea_logo.pdf} 
\caption{\label{fig2} HR diagram of the sample of O-type stars (filled red circles) in Trumpler 14, where the evolutionary tracks of various masses from Ekstr{\"o}m et al. 2012 are plotted (V/Vc=0).
}
\end{figure}

In the iDR4 we have 14 Carina early-type stars suitable for our analysis. Using automatized tools for the determination of stellar parameters (IACOB-GBAT \cite{ssimon12}) and FASTWIND stellar models \cite{sr97}, \cite{puls05}
we have obtained the fundamental parameters of all of them (rotational velocities, temperatures, gravities, He abundances and Q parameter).
In Figure~\ref{fig2} these stars are placed on a Hertzsprung-Russell diagram  where evolutionary tracks of stars of various masses from  \cite{ek13} are plotted. 3 of the stars are placed below the ZAMS, which could be due to a binary nature or a wrong distance determination. The new data provided by Gaia will help us to settle this point.

\section{FASTWIND vs CMFGEN \label{s13}} 

\begin{figure}[!h]
\center
%\par{
\includegraphics[width=9.5cm, angle=180]{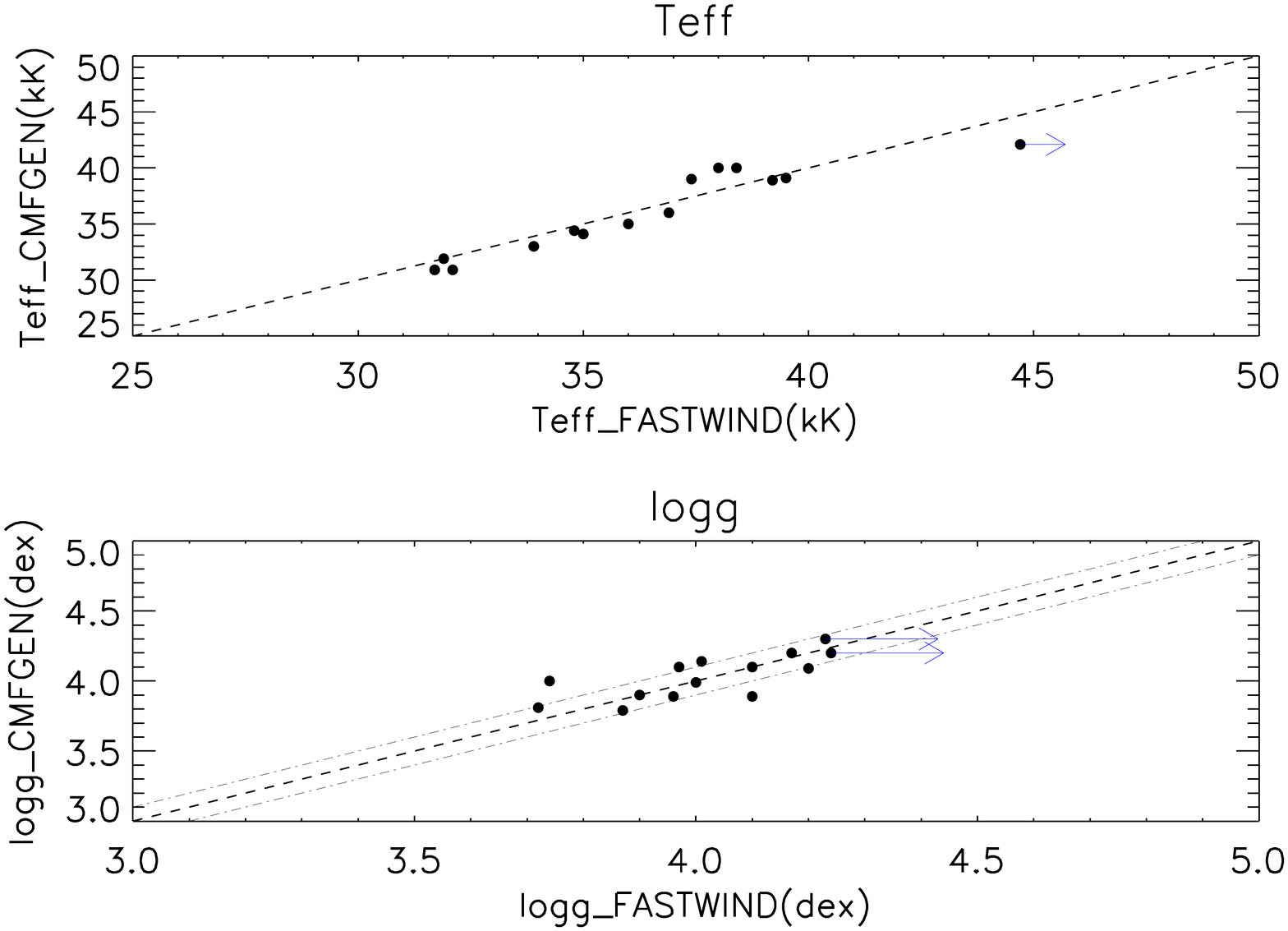}

%\par}
%\includegraphics[width=8.3cm,angle=0,clip=true]{sea_logo.pdf} 
\caption{\label{fig3} Comparison of effective temperatures and gravities obtained using FASTWIND and CMFGEN stellar atmospheres codes and considering the same vsini data. Blue arrows indicate lower limit values.
}
\end{figure}

It was found recently by other authors that gravities determined using FASTWIND were systematically lower by 0.12 dex compared to CMFGEN \cite{massey12}. In the GES WG-13 there are 2 groups analyzing O-type stars but using different stellar atmosphere codes: FASTWIND and CMFGEN, so we have compared results from both groups in order to confirm or reject this difference. Note however that, different to \cite{massey12}, our results have been obtained by two different groups. Taking into account that errors in the projected rotational velocity (vsini) affect the determination of logg \cite{sabin14}, we have analyzed the sample from Section 3 but using now the vsini considered by the CMFGEN group in order to compare only codes. We do not obtain a clear systematic effect on the gravity (see Figure~\ref{fig3}) and thus we can not confirm the results from \cite{massey12}. The new data of the imminent iDR5 will provide us with a bigger sample to corroborate it.

%
%
% Do not delete the next line
\small  % Do not delete
%
%%% Comment the following line if you do not have acknowledgments.
%\section*{Acknowledgments}   % Do not delete if you declare acknowledgments
%
%%% ACKNOWLEDGMENTS
%%% ACKNOWLEDGMENTS
%If you do not have any acknowledgments, you may comment this Section. We certainly acknowledge the Editors of Highlights of Spanish Astrophysics VII and VIII for their help while preparing this documentation.

%
% Do not delete the next few lines

%
\end{document}